\documentstyle[12pt,aaspp4]{article}

\def\msun{{\,M_\odot}} 
\def\gram{\hbox{g}}

\def\sec{\hbox{s}}

\def\days{\,\hbox{days}}

\def\yr{\hbox{yr}}

\def\km{\, \hbox{km}}
\def\kpc{\, \hbox{kpc}}
\def\pc{\, \hbox{pc}}

\def\Msun{M_\odot}
\def\bp{Paczy\'{n}ski}
\def\kms{\km / \sec}

\def\R0{R_0}
\def\rthat{\hat{r_t}}
\def\LE{{\cal E}}

\def\sech{\hbox{sech}}
\def\erfc{\hbox{erfc}}

\begin{document}
 
\title{ Microlensing of Globular Clusters as a Probe of Galactic Structure }
 
\author{James E. Rhoads}
\affil{Kitt Peak National Observatory, 950 North Cherry Avenue,
Tucson, AZ 85719; jrhoads@noao.edu}
\author{Sangeeta Malhotra}
\affil{Infrared Processing and Analysis Center, M.S. 100-22, CalTech,
Pasadena, CA 91125; san@ipac.caltech.edu}

\begin{abstract}

The spatial distribution of compact dark matter in our Galaxy can be
determined in a few years of monitoring Galactic globular clusters for
microlensing.  {Globular clusters are the only dense fields of
stars distributed throughout the three-dimensional halo} and hence are
uniquely suited to probe its structure. The microlensing optical
depths towards different clusters have varying contributions from the
thin disk, thick disk, bulge, and halo of the Galaxy. Although
measuring individual optical depths to all the clusters is a daunting
task, we show that interesting Galactic structure information can be
extracted with as few as $40$--$120$ events in total for the entire
globular cluster system (observable with 2--5 years of monitoring). The
globular cluster microlensing is particularly sensitive to the core
radius of the halo mass distribution and to the scale length, surface
mass density, and radial scale height variations of the thin disk.

\end{abstract}

\section{Introduction}

The search for compact dark matter in the mass range
$10^{-6}$--$10^{2} \msun$ using microlensing (\bp\ 1986) has come to
fruition with the detection of many microlensing events towards the
Galactic bulge, Large Magellanic Cloud, and Small Magellanic Cloud by
various projects to monitor millions of stars (Alcock et al. 1993,
1996, 1997a,b; Aubourg et al. 1993; Udalski et al. 1993). These studies have
confirmed the presence of a massive bar in the Galaxy by detecting
more microlensing events than expected towards the Galactic bulge
(Udalski et al. 1994; Alcock et al. 1995a; \bp\ et al. 1994; Zhao, Spergel,
\& Rich 1995) and have found evidence for a massive halo by detecting 8
times more events towards the Large Magellanic Cloud (LMC) than can be
explained by known stellar populations (Alcock et al. 1997a).
 
Uncertainties in the spatial distribution, kinematics, and 
masses of the lensing objects remain (cf. review by \bp\ 1997).  One
of the major uncertainties is our incomplete knowledge of Galactic
structure.  Possible locations for the lenses include a maximal
Galactic disk (Kuijken 1997), the LMC's disk (Sahu 1994), and the Galaxy's 
halo.  Here we propose using globular cluster stars as background
objects to discriminate the spatial distribution of the lensing
objects statistically. This is possible because the globular clusters
are distributed throughout the halo of our Galaxy, albeit concentrated
towards the center. Therefore we can probe many lines of sight through
varying amounts of the halo, disk, and bar of the Galaxy to determine
the relative contributions of each to microlensing.  Similar
strategies using the ratios of Bulge, LMC and SMC optical depths have
been suggested by Sackett and Gould (1993) to constrain halo
flattening, and by Gould, Miralda-Escud\'{e} \& Bahcall (1994) to
constrain relative contributions of the thin and thick disks and the
halo.

Another concern for microlensing studies is that the Galactic halo may
be clumpy, as would be expected under the Searle and Zinn (1978) picture
of Galaxy formation. The discoveries of the Sagittarius dwarf galaxy
(Ibata, Gilmore \& Irwin 1994) and of a possible intervening
population towards the LMC (Zaritsky \& Lin 1997) provide examples of
clumps of stars, in tidal tails or dwarf spheroids in the halo, that
might raise the number of microlensing events towards some lines of
sight (Zhao 1997). Monitoring microlensing events towards many
Galactic directions where globular clusters are located will average
over such fluctuations in halo density and thereby alleviate the
potential worry about non-representative lines of sight.

Because the globular clusters are distributed throughout the Galaxy's
halo, such a campaign of monitoring along different Galactic
directions would be invaluable for weighing the compact massive
component of the Galaxy (in stars or brown dwarfs) not only in the
halo, but also in the thick disk, thin disk, and bulge.  Measuring
individual optical depths to the clusters is not practical, but
interesting Galactic structure information can be extracted with as
few as $40$--$120$ events in total for the entire globular cluster
system (observable with 2--5 years of monitoring).

In section 2, we review necessary background information on
microlensing and on the globular clusters.  We take M~15 as an example
and discuss the observability of lensing events in globular clusters,
using resolved stars and pixel lensing. In section 3, we study the
number of events required to distinguish between representative Galaxy
models using Monte Carlo simulations.  Finally, in section 4 we
discuss some more implications of this work and summarize our
conclusions.  An appendix contains analytic results that supplement
the simulations in section 3.

\section{Microlensing of a Globular cluster}

The optical depth to microlensing is defined as the probability that a
background source will be amplified by a factor $1.34$ (Refsdal 1964;
Vietri \& Ostriker 1983).
This amplification corresponds to the background source lying inside the
Einstein radius of the lens.  Because the area in the Einstein ring is
proportional to lens mass, the optical depth depends only on the mean mass
density and not on the masses of individual lensing objects, provided only
that the Einstein radius substantially exceeds the physical size of both
source and lens.  This condition requires lensing masses $M \ga 10^{27}
\gram \sim M_\oplus$ for length scales typical in the Galaxy.  The geometry
of source, lens, and observer affect the Einstein radius, and the full
expression for the lensing optical depth $\tau$ is (Vietri \& Ostriker
1983)
\begin{equation}
\tau = \int_0^D { 4 \pi G \rho \over c^2 } L \left(1 - {L\over D}
\right) dL
\label{taudef}
\end{equation}
where $D$ is the distance to the background source, $L$ is the distance
along the line of sight, $\rho$ is the local mass density at distance
$L$, and $G$ and $c$ are the gravitational constant and the speed of
light.


\begin{figure}[htbp]
\epsfxsize=11.0cm\epsfbox{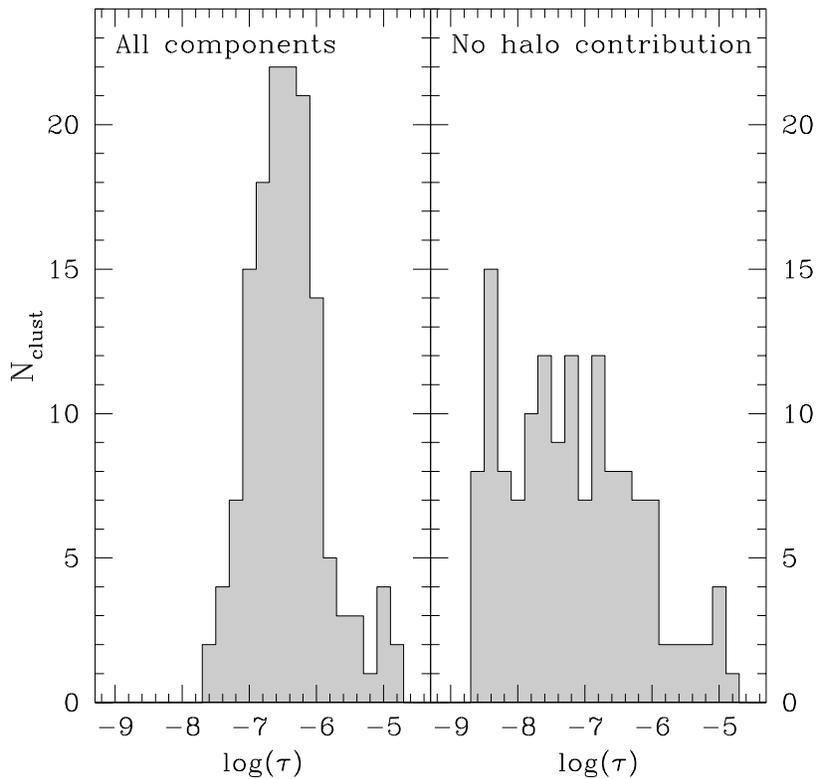}
\caption{
Histograms of microlensing optical depths to Galactic globular
clusters under a ``standard'' model (model 5,6,7a in table~1).  Left
panel: Histogram of total optical depths.  Right panel: Histogram of
the optical depth due to the disk and bar components alone.  The
latter case is a lower bound to the expected optical depths, and omits
the lensing contribution of both the stellar halo and the dark matter
halo.}
\end{figure}

The Galaxy has $\sim 150$ globular clusters, distributed in a spheroidal halo
about the Galactic center.  Each typically contains $\sim 10^4$--$10^5$
stars to a limiting magnitude of $V \le 22$, giving a total of $\simeq
10^6$--$10^7$ stars to monitor. The optical depth to the clusters is
$10^{-6}$--$10^{-8}$, depending on the cluster location in the Galaxy and the
Galactic model (Figure 1).  In addition to the lensing optical depth,
a full analysis would need to consider the mass distribution of the
lenses and their velocities in order to predict the event rate and
duration. These can then be compared to the time sampling of
monitoring projects to determine the efficiency of a search. Such
exhaustive analyses have been carried out for the Galactic bulge,
Magellanic clouds, and M31 (Alcock et al 1995b), which are already
being monitored.  Observationally, the median duration of events
towards the bulge is 30 days (Alcock et al 1997a).  The typical
timescale for globular cluster microlensing events may be comparable
or somewhat different, depending on the relative kinematics of
globular clusters and the lensing populations.

The number of stars in a globular cluster that can be resolved and
individually monitored from the ground will depend upon the number of
stars in the cluster, the degree of concentration, the distance to the
cluster and the seeing conditions. In the inner regions, crowding (up
to 130 stars per square arcsecond for $V < 22$) is too severe for
individual stars to be observed.  Here one can instead monitor the
brightness variations of pixels, which are a combination of many
stars; this is the so-called pixel-lensing regime (Crotts 1992; Colley
1995).

\label{pixlens}

We account for crowding by first calculating the effective number of
stars that can be monitored in the cluster M~15, and then scaling the
results for clusters of other luminosities and heliocentric distances.
We chose M~15 as an example because stars in the interior core (down
to $0.3 \arcsec$) have been resolved by Hubble Space Telescope
observations (Guhathakurta et al. 1996).  M~15 is at a distance of
$\simeq 10 \kpc$, about the median distance to globular clusters.  It
has a high concentration index, which will lead us to a conservative
(low) estimate for the fraction of resolved stars in a typical
cluster.  M~15 is also 2.7 times brighter (in V-band absolute flux)
than the average cluster.

The stars in the crowded interiors of Globular clusters can be
monitored for pixel lensing events, where one pixel brightens due to
the microlensing of one of the unresolved stars.  The rate of
detectable lensing events for each star is lower than in the uncrowded
regime, because for a fixed amplification of the total flux in a
pixel, a star needs to undergo a larger magnification, which depends
on the fraction of light in the pixel arising from that star.  This in
turn depends on the number of stars in the pixel and the brightness of
the star. Since the crowding varies continuously, decreasing from the
unresolved core to resolved stars in the exterior, we calculate the
mean effective optical depth for the entire cluster using a pixel
microlensing formalism. In the outer regions of the cluster, where the
surface density drops to $\la 1$ star per pixel, the lensing cross
sections approach the typical Einstein ring area.

Suppose a microlensing event is claimed for a fixed fractional
increase in the brightness of a pixel. Following Colley (1995),
\begin{equation}
\Delta m = -2.5 \times \log {{A q f_* + f_{pix}}\over{q f_* + f_{pix}}} ~~,
\end{equation}
where $\Delta m$ is the brightness change in magnitudes, $A$ is the
amplification, $q$ is the fraction of the amplified
star's light falling in the peak pixel of the point spread function,
$f_*$ is the flux of the star, and $f_{pix}$ is the flux due to other
sources in the pixel.
The maximum radius $u$ that produces amplification $A$ is
given by (Refsdal 1964)
\begin{equation}
A={{u^2+2}\over{u\sqrt{u^2+4}}}
\end{equation}
where $u=R/R_E$, and $u^2$ gives the area as a fraction of the area of
the Einstein ring area $\pi R_E^2$. If the star were resolved and the
only one contributing to the flux in that pixel, $u=1$ and
$A=3/\sqrt{5}$. Using the luminosity function of M~15 and its
surface brightness profile (Guhathakurta et al 1996;
Trager, King, \& Djorgovski 1995) we
calculate the loss of efficiency, which is the ratio of the area in
the lens plane required to brighten the star in a field of surface
brightness $S$ vs when it is alone. This is a function of the surface
brightness of the pixel and the brightness of the star. Summing over
the stellar luminosity function $f(m)$ for M~15 (Guhathakurta 1996) in V band,
\begin{equation}
\LE(r)= \left. \int_{m_1=12}^{m_2=22} f(m) u(m)^2 dm
 \right/  \int_{m_1=12}^{m_2=22} f(m) dm  ~~,
\end{equation}
where $u$ has an implicit dependence on the local surface brightness 
and hence on $r$.
The value of $\LE$ varies from $6.4\times 10^{-4}$ at a radial
distance of $0.1 \arcsec$ from the center and approaches 1 at $r
\simeq 500\arcsec$ when stellar density reaches 0.12 stars per
square arcsecond. With $3 \times 10^{5}$ stars to a radius of
$600\arcsec$ and a mean efficiency of $\langle \LE \rangle = 65.9\%$,
the cluster can be treated like a cluster with $2 \times 10^5$ fully resolved
stars. 

For other globular clusters, we retained the radial profile and
luminosity function models from M~15.  The total numbers of stars were
scaled relative to M~15 according to each cluster's V band luminosity.
The projected distribution on the sky (and hence crowding) and the 
number of stars above our $V<22$ cutoff were  adjusted for the
distances to the individual clusters.  The mean number of effectively
resolved stars inside a $600\arcsec$ radius was $5.7\times 10^4$ per
cluster.

\section{Distinguishing between Galactic models} \label{simulations}

We report here on a series of simulations to determine how well
microlensing observations of globular clusters can distinguish among
plausible Galactic models.  In each test, the microlensing optical
depth towards each cluster is calculated under two assumed Galactic
models, and the probability of inferring the correct model is
calculated for a given number of events. The results are then inverted
to calculate the number of microlensing events needed to infer the
correct Galactic model from the distribution of lensing events among
the clusters with 90\% confidence.

The models generally consisted of three mass components: The thin
disk, the halo, and the central bar.  A fourth component, the thick
disk, was added in one model. All the mass is assumed to be in compact
objects. The optical depth to microlensing is calculated using
equation~(\ref{taudef}), substituting the density
$\rho=\rho_{\hbox{disk}}+\rho_{\hbox{bar}}+\rho_{\hbox{halo}}+
\rho_{\hbox{thick disk}}$.

The models were required to obey three constraints.  (1) Rotation Curve:
The potential of the Galaxy consisting of these components was
required to fit the rotation curve in the inner Galaxy excluding the
bar region ($2.5 \kpc < R < R_0$) and a flat rotation curve with
circular speed $220 \kms$ in the outer Galaxy ($R_0 < R < 16.5 \kpc$)
(Knapp, Tremaine, \& Gunn 1978; Gunn, Knapp, \& Tremaine 1979; Fich,
Blitz, \& Stark 1989; Malhotra 1994, 1995).  $\R0$ is the Sun-Galactic
Center distance, and is taken to be $8.5 \kpc$ throughout this work.
(2) The second constraint is the local surface mass density of the
thin disk, $\Sigma_0$, which lies in the interval $40 \Msun / \pc^2
\la \Sigma_0 \la 80 \Msun / \pc^2$ (Kuijken \& Gilmore 1989, 1991;
Bahcall, Flynn, \& Gould 1992; Flynn \& Fuchs 1994).  We implement
this by either fixing the local surface mass density or adjusting
other parameters to ensure that the final value lies in the
observationally acceptable range.  This resolves some of the
degeneracy between the radial force due to the disk and that due to
the halo. (3) A third, implicit, constraint is the mass profile of the
thin disk. From gas dynamics it is demonstrated to be an exponential
with roughly the same scale length as the light (Knapp 1990; Malhotra
1995, 1994).

{\sl The Thin Disk:} The functional form of the disk density component(s) was
taken to be
\begin{equation}
\rho_d = \Sigma_0 / (2 z_1) \times \exp(-[r-R_0]/r_1) 
\times (\sech[z/z_1])^2 ~~,
\label{rhodisk}
\end{equation}
where $r_1$ is the exponential scale length, $z_1$ is the scale
height, and $(r,z)$ are Galactocentric cylindrical coordinates
(Spitzer 1942).  We took $z_1 = 0.20 \kpc$ at the Solar radius.  The
scale height was usually kept constant with radius.  Such constancy is
indicated by most studies of external galaxies (van der Kruit \&
Searle 1982), and is
consistent with the diffuse near infrared light of the Milky Way
(Spergel, Malhotra, \& Blitz 1996; Freudenreich 1996).  However, there
may be exceptions to this rule (de Grijs \& Peletier 1997), so in one
case [model 5B] the scale height was allowed to increase with radius
according to $z_1 \propto 1/\Sigma(r) \propto
\exp(r/r_1)$.  This flaring is the appropriate behavior for a
self-consistent isothermal thin disk with constant vertical velocity
dispersion and exponentially decreasing surface density.

{\sl Halo:} A power law halo was adopted, following Evans (1993,1994)
and Alcock et al (1995b):
\begin{equation}
\rho_h = { v_0^2 \over 4 \pi G q^2 } { (1 + 2 q^2) r_c^2 + r^2 + 
(2 - q^{-2}) z^2  \over  \left(r_c^2 + r^2 + z^2/q^2 \right)^2 } ~~,
\label{rhohalo}
\end{equation}
where $r_c$ is the core radius of the halo and $q$ the halo flattening
parameter.  Note that Alcock et al's parameter $\beta$ has been set to
$\beta \equiv 0$, so that the asymptotic behavior at large radii is a
flat rotation curve with circular speed $v_0$.
The power law halo component admits a
self-consistent distribution function and an analytic potential; the
latter was useful in ensuring that the models match observational
rotation curve constraints.
We do not include halo truncation, since relatively few globular
clusters lie at galactocentric radii $\gg 10 \kpc$ while present
evidence suggests the halo extends to $40$--$50 \kpc$ (Little \& Tremaine
1987).

{\sl Bar/Bulge:}  Detailed models of the central bar of our Galaxy
have now been constructed by several groups (Dwek et al 1995; Zhao et al
1995; Stanek et al 1997).  We chose to follow the
Zhao et al (1995) bar model:
\begin{equation}
\rho_b = { M_b  \over 8 \pi a b c } e^{ -s^2 / 2 }
\label{rhobar}
\end{equation}
where $a=1.49\kpc$, $b=0.58\kpc$, and $c=0.40\kpc$ are the major axis,
minor axis, and vertical scale lengths; $ s^4 = \left[ (x/a)^2 +
(y/b)^2 \right]^2 + (z/c)^4 $; and $(x,y,z)$ is a Cartesian coordinate
system aligned with the bar.  We took
the major axis of the bar to be at an angle $\phi$ with respect to the
Sun-Galactic center line, and took the plane $z=0$ to be the Galactic
plane (so that the bar is not tilted).  In the rotation curve fits, we
did not include data closer than $2.5 \kpc$ from the Galactic center,
because of complex non-circular motions in the bar, and approximated
the potential of the bar as a point source potential of the same mass.

{\sl Tests:} Seven tests were executed in all.  Each determined the
sensitivity of globular cluster microlensing observations to one
parameter by comparing two paired models having a range of that
parameter.  Other structural parameters were adjusted as needed to
maintain consistency with the rotation curve.  The models are
summarized in table~1.  Test~1 is designed to discriminate between
models with a large and small halo core radius.  Other components
necessarily become more massive in the case of a large core radius.
Test~2 varies the halo flattening parameter, while keeping everything
else fixed.  Test~3 uses a fixed disk mass but a range of disk scale
length; the halo parameters here change substantially also.  Test~4
primarily explores local disk surface mass density.  Test~5 allows the
scale height of the thin disk to increase with Galactocentric radius.
Test~6 includes a thick disk component of scale height $600\pc$,
surface density $1 \Msun/\pc^2$, and scale length identical to the
thin disk.  Finally, test~7 varies the angle of the Galactic bar with
respect to the Sun-Galactic center line.  Note that the rotation curve
constraint couples the parameters of different Galactic components in
our models.  Differences between microlensing optical depths in models A and B
of a test may thus be due to changes in more than one Galactic component.

For each model, the microlensing optical depth to each cluster was
determined by evaluating equation~(\ref{taudef}) numerically for the
density model given by equations~(\ref{rhodisk})--(\ref{rhobar}).  Our
cluster sample consisted of 
140 globular clusters whose coordinates, distances, 
and absolute V band magnitudes are tabulated in Djorgovski \& Meylan (1993)
and Djorgovski (1993).  As a consistency check, we also implemented
some of the halo models by Sackett \& Gould (1993), and verified that
our numerical integration code reproduced their optical depths towards
the Magellanic Clouds and Galactic Bulge within 1\%.

Given the optical depths, we then approximated the effect of an
observing campaign by defining the effective number $n$ of trials per
star monitored.  This is essentially the number of independent lensing
events we expect to observe per unit optical depth, or equivalently,
the number of event durations spanned by the observing campaign.
Formally, $n$ is defined by  $\langle E_* \rangle = n \tau$, where
$\langle E_* \rangle$ is the number of events expected per star
monitored and $\tau$ is the lensing optical depth.
Accounting for inefficiencies in the monitoring program, we write $n
\approx \epsilon t / d$.  Here $t$ is the duration of the monitoring
program, $d$ is the mean duration of lensing events, and $\epsilon$ is
the fraction of events that will be detected given the time sampling
of the observing program.  The observing efficiency $\epsilon$ can be
approximated as $\epsilon = \epsilon_1 \epsilon_2$, where $\epsilon_1$
is the fraction of the year a source can be monitored and $\epsilon_2$
is the fraction of events with durations longer than the time between
observations and shorter than $t$.  Note, though, that a precise
calculation of $n$ would require the full 6-dimensional phase space
distribution function of the lenses to determine $\epsilon$.  We use
the approximations $d \sim 30 \days$ and $\epsilon \sim 0.3$, so that
$n \sim 3.5 \times (t/\yr)$. 

A simulation consists of assuming a uniform number of trials $n$ for all
clusters, and for each cluster drawing a random number of ``observed''
events from a Poisson distribution with mean $\langle E \rangle = N_* n
\tau$ under an assumed Galaxy model. The likelihood of obtaining the 
resulting fake data set is then computed under this model and under an
alternate model.  The model with the larger likelihood is taken as the
inferred model.  Many such simulations are run to obtain the
probability of inferring the wrong model for each assumed model /
alternate model pair.  The test is then repeated for many values of $n$.
In the limit of infinite data ($n \rightarrow \infty$), the
probability of error goes to zero.

Table~1 contains values of the effective number of trials $n_{90}$
required to distinguish between the two models in each test at the
90\% confidence limit (i.e., with a 10\% chance of inferring the
incorrect model).  Because this probability need not be the same for
the two input models, two values of $n_{90}$ are tabulated for each
test.  To determine $n_{90}$, we interpolate between measured error
probabilities in simulations at a few values of $n$ using scaling
relations from appendix~\ref{analytic}:
$p(\hbox{error})=\erfc(h)/2$ where $h \propto \sqrt{n}$.  These
scaling relations can also be used to determine the values of $n$
needed to achieve other confidence levels in distinguishing between
models.

\begin{table}[htb]
\begin{tabular}{|c|ccccccc|cc|}
\cline{1-10}
Model & $\Sigma_0$ &  $r_1$ &
$v_0$  & $r_c$ & $q$ &
$M_b$ & $\phi$ & $\log_{10}[n_{90}]$ & $E_{tot,90}$ \\
\cline{1-10}
1A & 40 & 3.5 &         175 &{\bf 2.5} & 0.8 &  1.00 & 25  & 0.82 & 51.3\\
1B & 50 & 3.0 &         201 &{\bf 10.0} & 0.8 & 1.31 & 25  & 0.78 & 42.5\\
2A & 50 & 3.0 &         183 & 6.9 & {\bf 1.0} & 1.00 & 25  & 1.60 & 263 \\
2B & 50 & 3.0 &         183 & 6.9 & {\bf 0.625}&1.00 & 25  & 1.63 & 315 \\
3A & 67 & {\bf 3.5} &   161 & 5.43 & 0.8 &      1.00 & 25 & 1.10 & 89.9 \\
3B & 50 & {\bf 2.5} &   220 & 16.15 & 0.8 &     1.00 & 25 & 1.06 & 80.6 \\
4A & {\bf 80} & 3.31 &  200 & 15.21 & 0.8 &     1.00 & 25 & 1.19 & 110 \\
4B & {\bf 40} & 2.65 &  200 & 8.6   & 0.8 &     1.00 & 25 & 1.21 & 108 \\
5A & 50 & 3.0 &         183 & 6.9  & 0.8 &      1.00 & 25 & 0.40 & 17.5 \\
5B$^*$ & 50 & 3.0 &         183 & 6.9  & 0.8 &      1.00 & 25 & 0.46 & 18.2\\
6A & 50 & 3.0 &         183 & 6.9  & 0.8 &      1.00 & 25 & 2.93 & 5884 \\
6B$^{**}$ & 49 & 3.0 &         183 & 6.9  & 0.8 & 1.00 & 25 & 2.93 & 5997 \\
7A & 50 & 3.0 &         183 & 6.9  & 0.8 &      1.00 & {\bf 25} & 2.22 &1155\\
7B & 50 & 3.0 &         183 & 6.9  & 0.8 &      1.00 & {\bf 12.5}&2.22 &1221\\
\cline{1-10}
\end{tabular}

\caption{Parameters for tested Galaxy models.  See text for
definitions of model parameters.
The units are as follows: $\Sigma_0$ ($\Msun/\pc^2$); $r_1$ ($\kpc$);
$v_0$ ($\kms$); $r_c$ ($\kpc$); $M_b$ ($10^{10} \Msun$); $\phi$
(degrees).  Remaining quantities are dimensionless.
The primary parameter tested in each model pair is in {\bf
bold face} in the table.  The thin disk vertical scale height 
$z_1 = 0.20 \kpc$ unless otherwise noted.
$\lg[n_{90}]$ and $E_{tot,90}$ give the results of the simulations.
$E_{tot,90}$ is essentially the number of lensing events required to
distinguish a model from its alternative at the $90\%$ confidence
level; $n_{90} = E_{tot,90} / (\sum N_{*,j} \tau_j)$ is the corresponding
effective number of trials (as defined in text).  The $1\sigma$ random error
in $n_{90}$ and $E_{tot,90}$ is $2.8\%$.
}
\tablenotetext{*}{Disk scale height increases with Galactocentric
radius.  See text for description.}
\tablenotetext{**}{Thick disk present.
Column density $1 \Msun/\pc^2$ at Sun; scale
height $0.60 \kpc$, scale length $3 \kpc$.}
\end{table}

These calculations do not include the optical depth due to
self-lensing (i.e., lensing of one cluster star by another).  This is
justified by the low geometric weight given to the integrand in
equation~\ref{taudef} when $D-L \sim 1 \pc$.  Dimensional analysis
shows that for a cluster having mass $M_{cl}$ and characteristic
radius $a$, the self-lensing optical depth should be of order $G
M_{cl} / (a c^2)$.  We did more detailed calculations for clusters with
density $\rho \propto \left(1 + (r/a)^2 \right)^{-1}$ for $r < r_t = a
\rthat$ and $\rho \equiv 0$ for $r>r_t$.  These yield mean optical depth to
self-lensing $\langle \tau_s \rangle
= G M_{cl} f(\rthat) / (2 a c^2) = 2.4 \times 10^{-9} (M_{cl} / 10^6
\Msun) (1 \pc / a) f(\rthat)$, where $f \approx 3.5/\rthat$ to $15\%$
accuracy for $5 \la \rthat \la 500$.  For plausible cluster
parameters, then, $\langle \tau_s \rangle \la 10^{-9}$, which is
negligible compared to foreground optical depths (Figure 1).

The microlensing observations of the globular cluster system are most
powerful in our tests 5 (flat vs. flaring disk), 1 (halo core radius),
3 (disk scale length), and 4 (local disk surface density).  In
general, the microlensing optical depths towards the globular clusters
are sensitive to the large scale structural parameters of the Galaxy's
most massive components, while they are not very sensitive to the
presence or absence of a low mass thick disk (test 6) or to changes in
the bar geometry within plausible limits (test 7).  The sensitivity to
test~5 is in part because model 5B (disk scale height increasing with
galactocentric radius) is an extreme flaring model.  We therefore take
test~1 as our most favorable result for fully realistic models.

Our simulations were constructed on the assumption that the same
observational program is applied to each globular cluster.  This
assumption could be replaced by assigning different priorities to
different clusters.  The most reasonable observational program might
leave out the poorest clusters entirely as providing an
insufficient number of stars to be useful, while selectively including
or rejecting other clusters depending on what facet of Galactic
structure is being probed.  For example, tests of disk flaring might
omit observations of clusters at high Galactic latitude, where the
disk contribution to $\tau$ is negligible.  Illustrative simulations
with restricted subsets of clusters show that the duration of a
monitoring campaign need only be doubled to achieve $90\%$ confidence
discrimination between model pairs with a well chosen sample of $\sim
10$ clusters.  This strategy allows a specific model pair to be tested
using a small fraction of the telescope time needed for a microlensing
survey of the entire cluster system.  Its results will, however, be
more strongly affected by halo substructure.  In addition, the optimum
cluster subsample may be different for different model pairs.

\section{Conclusions}
From the simulations in this paper we see that the total number of
events required to distinguish between models is typically $\sim
40$--$120$ for the whole set of clusters. 

The precise observing time required to achieve $\sim 100$ total events
also depends on the lensing event duration. In this paper we have
simply adopted a duration of 30 days, which is the median duration of
microlensing events seen so far. The actual event durations will
depend upon the masses and the spatial distribution of the lenses, and on the
relative kinematics of the lens and the globular clusters in question.
Statistics of event durations will, of course, provide further
constraints on the masses, kinematics and spatial distribution of the
lenses. With this event duration of 30 days and monitoring the
globular cluster system to a magnitude limit of $V< 22$ one should be
able to distinguish between typical Galactic models discussed here
[tests 1, 3, and 4] in 2--5 years. For comparison, ongoing
microlensing projects have found 45 events in a year after monitoring
$\simeq 10^7$ stars towards the bulge and 8 events for $\simeq 10^7$
stars towards LMC in 2 years. The event rate towards the
globular cluster system should be between these two rates, because the clusters
are distributed over a range of Galactic coordinates with a concentration 
towards the Galactic center.  Two to five years is not
an excessive amount of time given that the MACHO, OGLE, and EROS
monitoring programs have been in operation for 5 years (1992-1997).

\appendix
\section{Analytic results for comparison of two models}
\label{analytic}
Consider a comparison of two models under a fixed observing strategy.
Let $\mu_{j:i}$ be the expected number of events and $\tau_{j:i}$ the
optical depth towards cluster $j$ under model $i$.  The two are related
by $\mu_{j:i} = n_j N_{*,j} \tau_{j:i}$, where $n_j$ characterizes the
observing time spent on cluster $j$ (section~\ref{simulations})
and $N_{*,j}$ is the effective number of resolved stars in cluster $j$
(section~\ref{pixlens}).
Then if model~1 is correct, the probability of obtaining a data set whose
likelihood under model~2 exceeds that under model~1 is
\begin{eqnarray}
\nonumber P(\hbox{error})  & = &
{1 \over 2 } - {1 \over 2 \pi} \int_{-\infty}^{\infty} {1 \over k}
\sin\left\{ \sum_{j=1}^{N_c} k \left( \mu_{j:2} - \mu_{j:1} \right) +
\mu_{j:1} \sin\left[ k \ln\left( \mu_{j:1}/\mu_{j:2} \right) \right]
\right\} \\
 & &  \times \exp \left\{  \sum_{j=1}^{N_c} \mu_{j:1} \left(
\cos\left[ k \ln\left(  \mu_{j:1}/\mu_{j:2} \right) \right] - 1
\right) \right\} dk  ~~ .
\end{eqnarray}
A more readily calculated (though approximate) scaling is that
$ P(\hbox{error}) \approx \erfc\left( h \right) / 2$, 
where $\erfc(x) \equiv 2 \int_x^\infty e^{-x^2} dx / \sqrt{\pi}$
is the complementary error function, and where multiplying all the $n_j$ by a
factor $\alpha$ changes $h$ by a factor $\sqrt{\alpha}$.
Both these results are consistent with our simulations over the range of
$P(\hbox{error})$ investigated.  Interested readers may contact the
authors for derivations of these results.

\acknowledgements
We thank Jill Knapp, Ken Mighell, Bohdan \bp\ and David Spergel for
helpful discussions.  JER's work is supported by a Kitt Peak
Postdoctoral Fellowship.  Kitt Peak National Observatory is part of
the National Optical Astronomy Observatories, operated by the
Association of Universities for Research in Astronomy.

\end{document}